\documentclass[12pt]{article}
\usepackage[T1]{fontenc}
\usepackage[utf8]{inputenc}
\usepackage[margin=1in]{geometry}
\usepackage[affil-it]{authblk} 
\usepackage{etoolbox}
\usepackage{lmodern}
\usepackage{cite}
\usepackage{changes}
\usepackage{microtype}
\usepackage{booktabs}
\usepackage{amssymb}
\usepackage{float}
\DeclareUnicodeCharacter{030C}{\v{c}}
\usepackage{url}
\usepackage{xcolor}

\title{Utilizing photonic band gap in triangular silicon carbide structures for efficient quantum nanophotonic hardware}

\def\correspondingauthor{\footnote{Corresponding author: prsaha@ucdavis.edu}}

\author[1]{Pranta Saha\correspondingauthor{}}
\author[1]{Sridhar Majety}
\author[1]{Marina Radulaski}
\affil[1]{Department of Electrical and Computer Engineering, University of California, Davis, CA 95616, USA}

\date{\vspace{-2em}}

\usepackage{graphicx}
\usepackage{amsmath}
\usepackage{braket}
\usepackage{titlesec}
\usepackage{siunitx}
\usepackage[version=4]{mhchem}
\usepackage{multirow}
\begin{document}

\maketitle
\begin{abstract}
Silicon carbide is among the leading quantum information material platforms due to the long spin coherence and single-photon emitting properties of its color center defects. Applications of silicon carbide in quantum networking, computing, and sensing rely on the efficient collection of color center emission into a single optical mode. Recent hardware development in this platform has focused on angle-etching processes that preserve emitter properties and produce triangularly shaped devices. However, little is known about the light propagation in this geometry. We explore the formation of photonic band gap in structures with a triangular cross-section, which can be used as a guiding principle in developing efficient quantum nanophotonic hardware in silicon carbide. Furthermore, we propose applications in three areas: the TE-pass filter, the TM-pass filter, and the highly reflective photonic crystal mirror, which can be utilized for efficient collection and propagating mode selection of light emission.
\end{abstract}

\section{Introduction}

Color centers are defects in wide band gap single-crystal materials that can emit single-photons and spin-entangled photons 
which act as quantum information carriers. Silicon carbide (SiC) is one of the most notable quantum hardware platforms since it hosts a collection of optically addressable color centers \cite{norman2021novel} with long spin coherence times \cite{widmann2015coherent, christle2015isolated, seo2016quantum, babin2022fabrication}, excellent brightness \cite{castelletto2014silicon}, nuclear spins \cite{falk2015optical, klimov2015quantum}, and telecommunication wavelength emissions \cite{norman2021novel, majety2022JAP}, which are suitable properties for quantum information processing. On top of that, SiC has a large bandgap, high thermal conductivity, strong second-order nonlinearity, mechanical stability, and mature industrial presence \cite{song2011demonstration,radulaski2013photonic} making it a reliable platform for a variety of applications. Recently, photonics in triangular geometry has come into focus for increasing the efficiency of such solid-state quantum emitter processes \cite{burek2012free, song2018high, babin2022fabrication, majety2022JAP}. Triangular cross-section waveguide results from a bulk nanofabrication process called the angle-etch method that has been successfully implemented in both diamond \cite{burek2012free, atikian2017freestanding} and SiC \cite{song2018high, babin2022fabrication}. Previous fabrication processes were challenged by various imperfections that deteriorated the optical properties of the color centers or limited the robustness of the nanophotonic devices \cite{majety2022JAP}. On the other hand, triangular geometry offers emitter implantation in bulk substrates (free-standing waveguides), which ensures high-quality color centers with better coupling and can pave the way for efficient quantum photonic hardware.

Advancement of quantum information technology greatly depends on the realization of robust quantum networks \cite{ruf2021quantum, majety2022JAP,castelletto2022silicon} and generation of arbitrary all-photonic cluster states \cite{lindner2009proposal,buterakos2017deterministic,russo2019generation} which, in color center platforms,  are limited by the low photon collection efficiency. Color centers can have both transverse electric (TE) and transverse magnetic (TM) optical dipole-like emissions with a solid angle covering 4$\pi$. Hence, it is important to understand the TE/TM  dispersion relations, in the triangular waveguide geometry, with a view to controlling and steering the quantum light emitted from the color center by PBG formation for higher collection efficiency. 

The formation of photonic band gaps (PBGs) in photonic crystals (PhCs) has been explored in the past three decades after the discovery made by Yablonovitch and John \cite{yablonovitch1987PhC, john1987strong}. Although wave propagation in periodic structures has almost been a century-long study \cite{born1946wave}, PhCs have gained attention due to their robust light confinement capability, scalability, and small footprint \cite{shakoor2014Fp1,zhou2021Fp2}. Combination of different scatterers with unique lattice geometries \cite{villeneuve1992photonic, wang2001effects, johnson2001new, notomi2008ultrahigh, kalra2008modelling, zhou2010facile, song2011demonstration, quan2011deterministic, eguchi2012single, radulaski2013photonic, shi2017honeycombquasi_vff} has led to wider PBGs by reducing the structure symmetry and found applications in polarization beam splitters \cite{hou2015PBS1, butt2021PBS2}, optical logic gates \cite{anagha2020OLG1,swarnakar2020OLG2}, mirrors \cite{zhao2011mirror1,zhong2015mirror2}, sensors \cite{scullion2011sensing1,zhuo2014sensing2}, lasers \cite{painter1999laser1,lu2019laser2}, solar cells \cite{bermel2007solarcell, hwang2013solarcell2}, and more. Nevertheless, most of these studies have been conducted on either slab, rectangular, or cylindrical geometry. On the other hand, triangular cross-section PhCs have mostly been studied for constructing active photonic devices \cite{burek2014high, song2018high, majety2021Jphotonics} whereas the dispersion relations and PBG formations are yet to be discussed in detail. We explore these properties to advance the photonic integration in SiC color center based quantum devices.

In this paper, we begin with defining the relevant parameters for analyzing PBG formation in SiC triangular cross-section PhCs. Using the plane wave expansion (PWE) method, we calculate the band structures and discuss the dispersion relations related to individual geometry. We then observe the effects of parameter variation on PBGs and examine the designs in terms of nanofabrication. We conclude by proposing three photonic devices along with their structural configurations and  operational wavelength ranges which have the potential to be essential components of integrated photonic circuits.

\section{Triangular cross-section photonic crystal}
\begin{figure}[ht]
\centering\includegraphics [width=\linewidth]{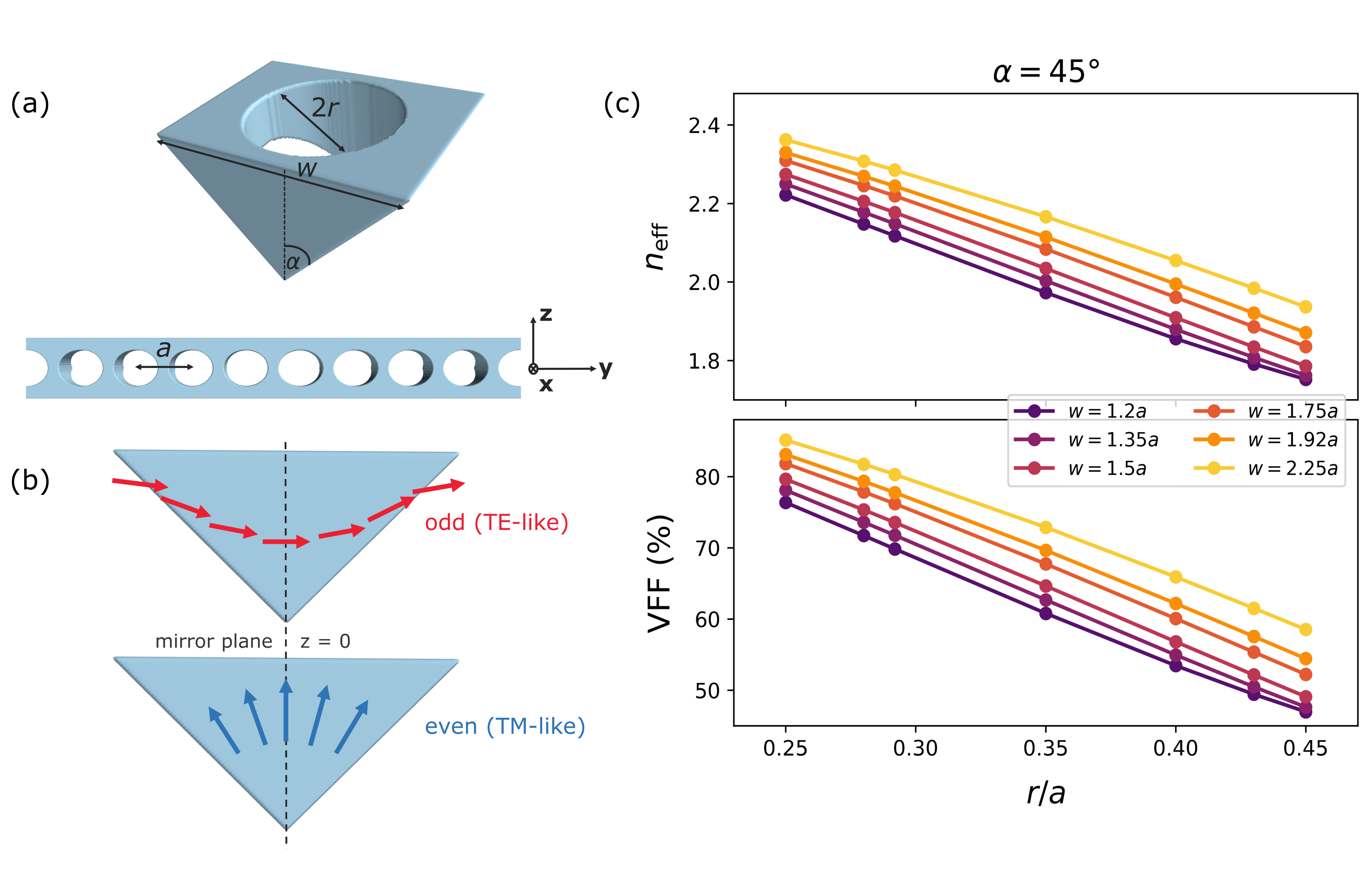}
\caption{(a) Schematic of the 1D photonic crystal unit cell. (b) Electric field lines (arrow) associated with corresponding TE-like (red) and TM-like (blue) modes. (c) $n_\mathrm{eff}$ and VFF calculation as a function of $r/a$ for $\alpha = 45^{\circ}$.}
\label{fig:fig_1}
\end{figure}

In this section, we define the parameters of the triangular cross-section photonic crystal. Traditionally, the periodic dielectric waveguides have periodicity along the direction of light propagation \cite{fan1995guided}. The triangular cross-section PhC in this study is realized in a similar fashion. Our 1D PhC structure is designed by inserting cylindrical air holes along $y$ axis in the SiC triangular cross-section waveguide as shown in Figure \ref{fig:fig_1}a. The most significant parameters of the PhC are its lattice constant $a$, waveguide width $w$, hole radius $r$, and etch angle $\alpha$. We examine three $\alpha$ values $35^{\circ}$, $45^{\circ}$, and $60^{\circ}$, which fall under realistic fabrication parameters of the state-of-the-art processes \cite{babin2022fabrication, song2018high, cranwell2021fabrication}. We vary the width $w$ from $1.2a$ to $2.25a$, and the radius $r$ from $0.25a$ to $0.45a$. We consider the refractive index of SiC to be $n_\mathrm{SiC} = 2.6$.

Based on the existence of a mirror symmetry plane ($z = 0$) perpendicular to the direction of periodicity, the photonic modes can be decoupled into TE-like and TM-like polarizations as illustrated in Figure \ref{fig:fig_1}b. Modes with electric field lines having odd symmetry about $z = 0$ are TE-like as the electric field lies in the plane of propagation. On the other hand, TM-like modes have even symmetry at $z = 0$ and the electric field lies in a direction perpendicular to propagation. 

We use the effective refractive index ($n_\mathrm{eff}$) as a useful parameter for understanding the modal profiles in a photonic device \cite{chen2015neff1, shi2017honeycombquasi_vff, majety2021Jphotonics}. It is defined as the ratio of propagation constant $(\beta)$ of a mode to the vacuum wavenumber $(2\pi/\lambda)$. In triangular geometry, the TE/TM polarized modes supported by the structure propagate according to their corresponding $n_\mathrm{eff}$ values. Modes with lower $n_\mathrm{eff}$ are not well contained within the structure and become evanescent \cite{majety2021Jphotonics}. As guided modes depend on $n_\mathrm{eff}$, which is a strong function of the effective dielectric present in the photonic device, its value can help us interpret the dispersion relations and changes in PBGs due to parameter variation in the proposed 1D PhC. Hence, in the following, we come up with an analytical expression for estimated $n_\mathrm{eff}$ derived from the volumetric fill factor (VFF) of SiC in the PhC structure:
\begin{equation}
    n_\mathrm{eff} = \mathrm{VFF} \times n_\mathrm{SiC} + (1 - \mathrm{VFF}) \times n_\mathrm{air} 
\end{equation}


Figure \ref{fig:fig_1}c demonstrates the changes in VFF and $n_\mathrm{eff}$ as a function of the normalized hole radii ($r/a$) in the $45^{\circ}$ angle-etched waveguide with various $w$ values. The plots show that $n_\mathrm{eff}$ reduces for higher $r/a$ and increases for higher $w$.  The latter happens due to the enlargement of the triangular cross-section with incremental $w$ which leads to greater $n_\mathrm{eff}$ and more supported modes. We observe similar trends and values for $35^{\circ}$ and $60^{\circ}$. 

\section{Methods}
\subsection{Dispersion relations in triangular geometry}

\begin{figure}[ht]
\centering\includegraphics [width=\linewidth]{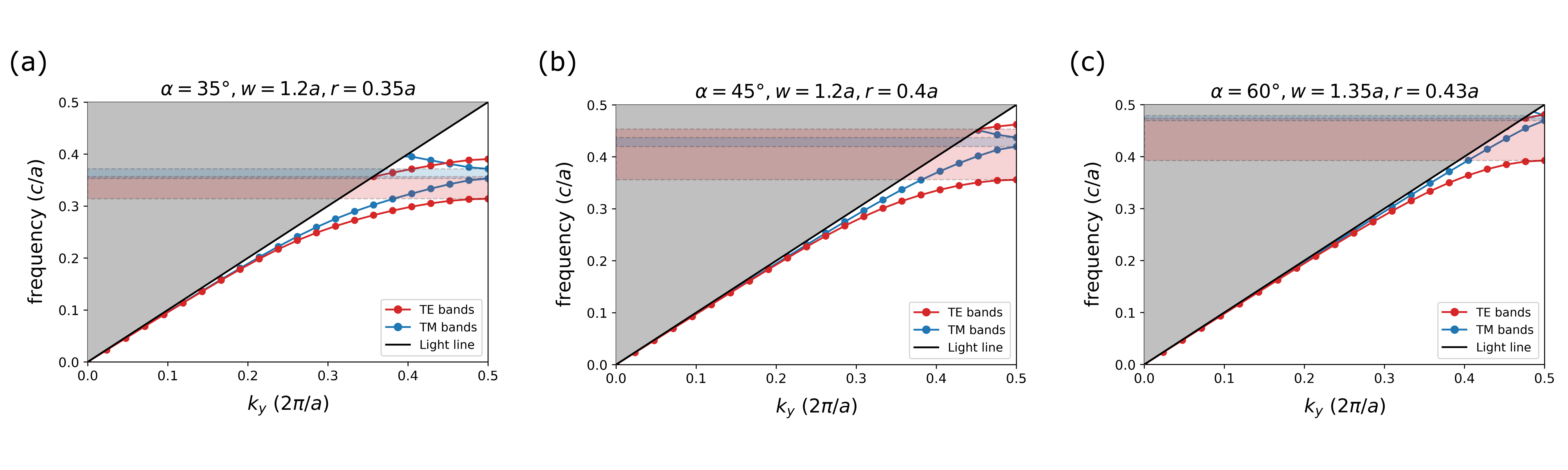}
\caption{Dispersion relations for TE (red) and TM (blue) modes in the triangular cross-section 1D photonic crystal. The red (blue) shaded regions show the photonic band gaps for the TE (TM) modes. Parameters of the photonic crystal are: (a) $\alpha = 35^{\circ}, w = 1.2a, r = 0.35a$. (b) $\alpha = 45^{\circ}, w = 1.2a, r = 0.4a$. (c) $\alpha = 60^{\circ}, w = 1.35a, r = 0.43a$.}
\label{fig:fig_2}
\end{figure}

While extensively studied in rectangular cross-section photonic crystals, the dispersion relations and PBG formations are not well understood in triangular geometry. Plane wave expansion (PWE) method has been widely used for analyzing PhCs due to its efficiency and accuracy in computing PBGs \cite{shi2004PWE1, quan2011deterministic, panda2021PhCPWE}. It is a direct frequency eigensolver method, derived from Maxwell's equations for a sourceless medium, where the eigenvalues are mode frequencies, and the eigenstates (plane wave solutions) are characterized by wavevector $\mathbf{k}$ and a band number. The irreducible Brillouin zone, which contains allowed wavevectors with non-redundant mode frequencies, lies in the range of $(0, 0, 0)$ to $(0, \pi/a, 0)$ in the $\mathbf{k}$-space for 1D PhC according to our definition of periodicity \cite{Joannopoulos:08:Book}. We have used MIT Photonic Bands (MPB) \cite{johnson2001mpb} to employ the PWE method for investigating band structure and PBG formation in the above mentioned irreducible Brillouin zone of the triangular cross-section PhC.  

The dispersion relations for three different PhC parameter sets ($\alpha, w, r$) are presented in Figure \ref{fig:fig_2}. In the band structure, the first TE/TM band (dielectric band) is the fundamental mode with the fewest nodes and the lowest frequency. The fundamental mode is well guided by the structure in all three PhCs. However, the higher-order mode (air band) for TE/TM is not entirely guided due to the light line effect, and becomes more radiative with $\alpha$ getting larger, as depicted in Figure \ref{fig:fig_2}. The TE (TM) PBG is formed by the range of frequencies present between the minima of the air band and the maxima of the dielectric band for the corresponding TE (TM) mode below the light line $(\omega < ck)$. All three angle-etched 1D PhCs show both TE and TM band gaps. TE band gaps are larger than TM gaps due to the connected dielectric lattice structure which is in accordance with the general intuition \cite{joannopoulos1997twist}. Figure \ref{fig:fig_2} also shows that $45^{\circ}$ and $60^{\circ}$ PhCs exhibit comparable TE gaps, while the $35^{\circ}$ band gap is reduced. On the other hand, the TM gaps are comparable in $35^{\circ}$ and $45^{\circ}$ cases, and reduced for  $60^{\circ}$.

Complete PBG refers to the region of forbidden propagation frequencies in the band structure regardless of polarization and is formed in the overlap of the TE and TM band gaps. Though conventional multilayered 1D PhCs lack a complete PBG \cite{butt2021PhCreview}, the triangular cross-section geometry offers complete PBG for all three studied angles $\alpha$. As PBG is the principal characteristic of a PhC, it is desired to obtain a design with as wide complete PBG as possible. Figure \ref{fig:fig_2} demonstrates that the largest complete PBG can be achieved with the $45^\circ$ structure as the TM gap is completely buried within the TE gap. Complete PBG in the other two cross-sections occurs from a small overlapping region between the TE and TM band gaps. Even though in two $(w, r)$ sets, the $60^\circ$ geometry shows buried complete PBGs, these gaps are about four times narrower compared to the $45^\circ$ case. Therefore, $45^\circ$ angle-etched triangular cross-section 1D PhC is more favorable for polarization-independent light confinement.

\subsection{Effects of parameters on PBG formation for $\boldsymbol{\alpha = 45^{\circ}}$}

\begin{figure}[ht!]
\centering\includegraphics [width=0.8\linewidth]{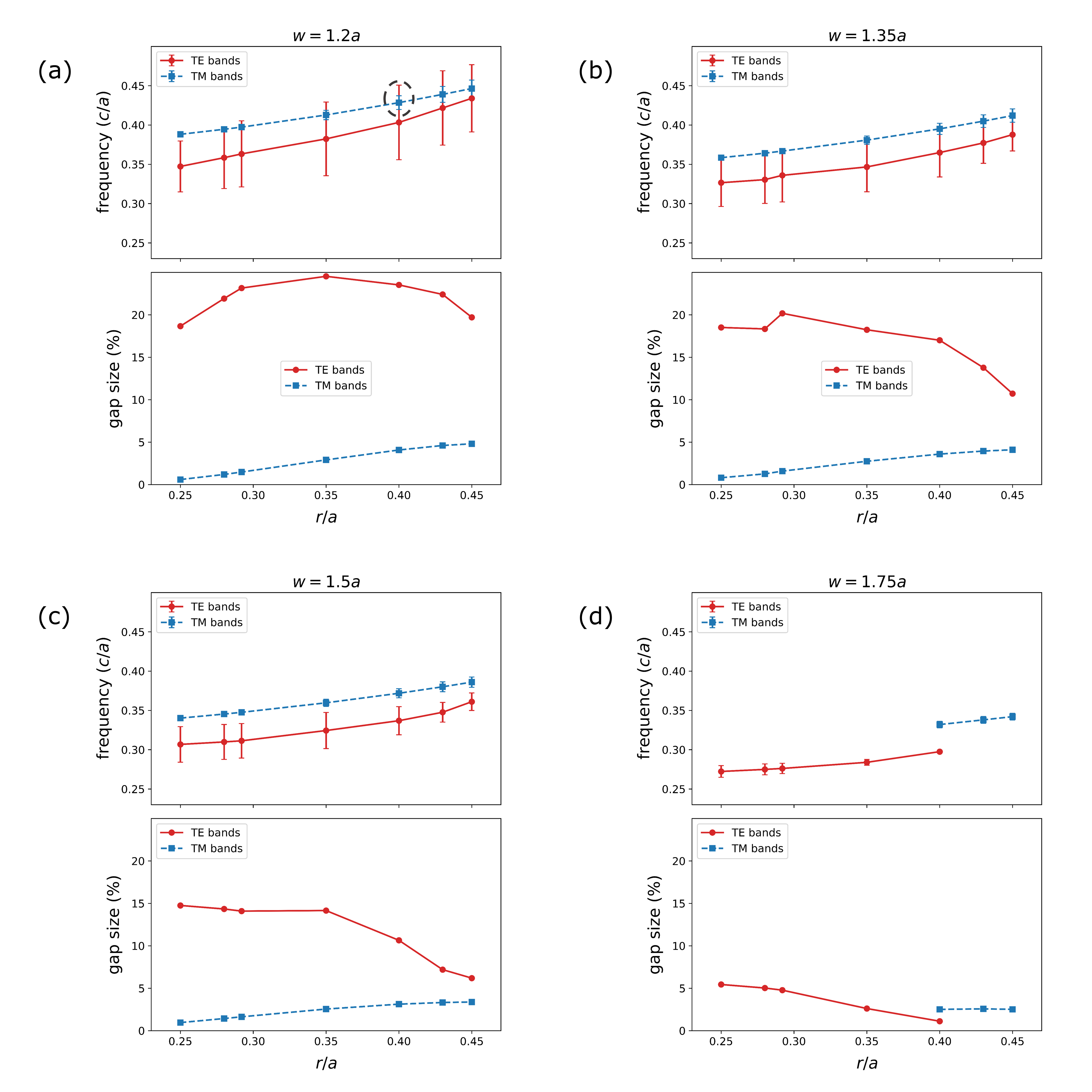}
\caption{(a)-(d) TE/TM photonic band gap in frequency ($c/a$) and gap size ($\%$) w.r.t. normalized hole radii $(r/a)$ for the $45^\circ$ triangular cross-section waveguide with $w$ values of $1.2a, 1.35a, 1.5a$, and $1.75a$, respectively. In the frequency plots, the center points are midgap ($f_\mathrm{m}$) frequencies and the error bars indicate band widths ($\Delta f$) for the corresponding TE/TM modes. The dashed circle shows the TE-TM gap overlap for the dispersion relations demonstrated in Figure \ref{fig:fig_2}b.}
\label{fig:fig_3}
\end{figure}

In this section, we further analyze the effects of parameters to achieve the best performing design in the $45^\circ$ angle-etched waveguide. Scale-invariant nature of Maxwell's equations actuates the idea of presenting parameters and results in terms of lattice constant $a$. Consequently, the gap width $\Delta f$, where $f$ is frequency expressed in units of $c/a$, is not a useful measure to understand the extent of a PBG. The gap to midgap ratio ($\Delta f/ f_\mathrm{m}$), popularly known as the gap size, is a more telling characterization of the gap width as it is independent of the scaling. Figure \ref{fig:fig_3} manifests TE-TM gap ($c/a$) and gap size ($\%$) variation with $r/a$ in $\alpha = 45^\circ$ waveguides having several widths $w$. With incremental $r/a$, the $n_\mathrm{eff}$ decreases leading to an increase in $f_\mathrm{m}$ for both the TE and TM band gaps, consistent with the literature \cite{anderson2017model}. The opposite happens when $w$ increases, due to the increased $n_\mathrm{eff}$, for a corresponding $r/a$ value. 

In smaller $w$ such as $1.2a$ and $1.35a$, the TE gap size initially increases with $r/a$ owing to fewer supported modes as a result of lower VFF and $n_\mathrm{eff}$, but shrinks after reaching the resonant condition at which the gap size is maximum. This happens on account of the reduction in effective dielectric contrast with higher $r/a$ values. Apparently, TE gap size in PhCs with larger widths $w$ is smaller due to greater $n_\mathrm{eff}$, and TE band gap vanishes above $w = 1.75a$. On the contrary, the TM gap size monotonically grows with $r/a$, however, the band gap disappears for larger widths, identical to the TE case. From Figure \ref{fig:fig_3} and the above discussion, it is evident that complete PBG (either buried or overlap) mostly occurs for smaller widths and TE/TM band gap totally vanishes for waveguides with larger widths ($w > 1.75a$).

\section{Results and discussion}
\begin{figure}[ht]
\centering\includegraphics [width=0.6\linewidth]{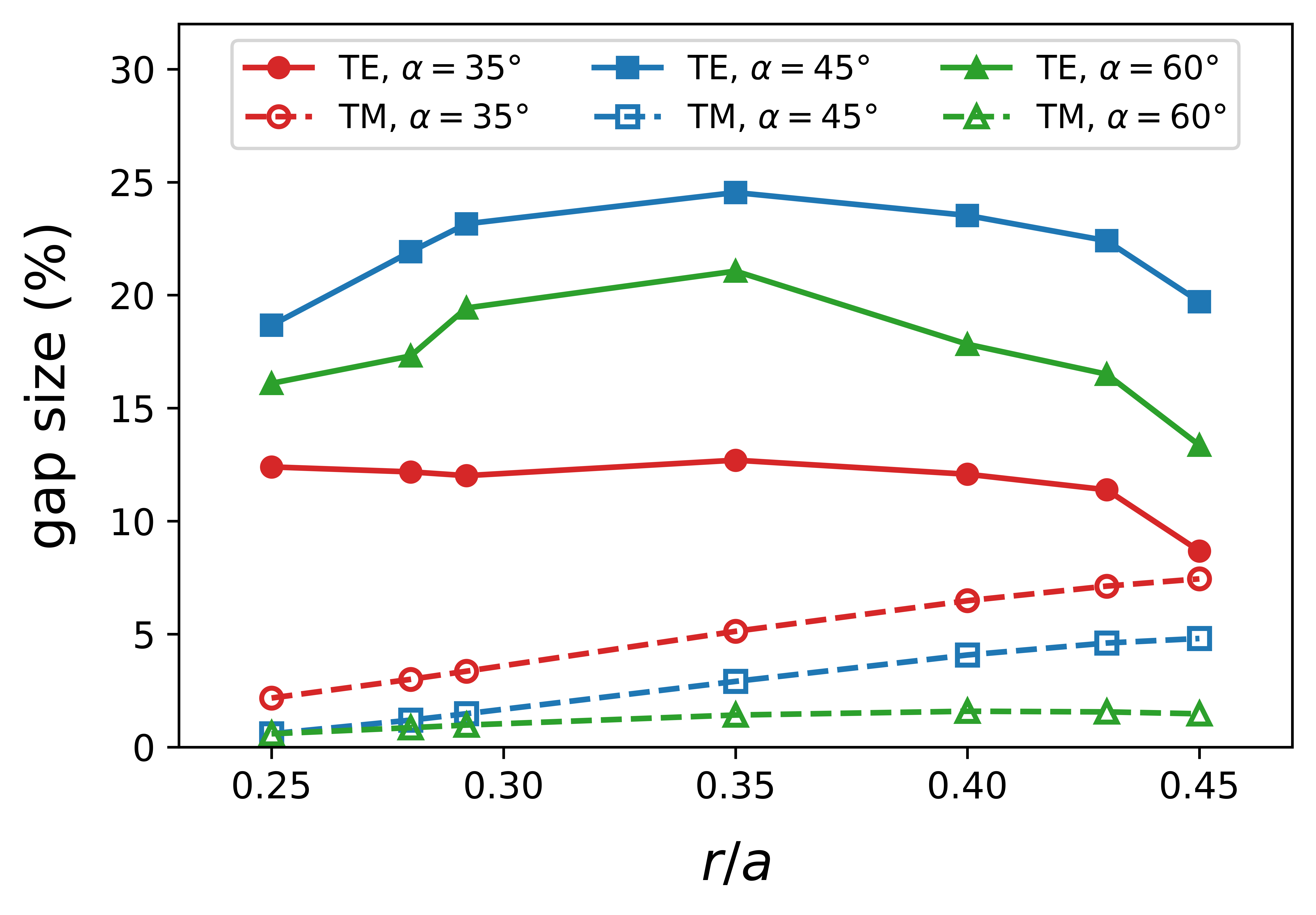}
\caption{TE/TM gap size ($\%$) in $w = 1.2a$ with varying normalized hole radii $(r/a)$ for three $\alpha$ values.}
\label{fig:fig_4}
\end{figure}

The demonstrated work provides insights into the dispersion relations in the non-standard, triangular, geometry of photonic crystals. Figure \ref{fig:fig_4} delineates a general comparison of TE/TM gap sizes among three etch angles for a constant width $(w = 1.2a)$. We observe that unique trend emerges from unique geometry. For the $35^\circ$ triangular cross-section, the TE gap size appears stable with changes in hole radii, whereas the TM gap size variation is identical to the $45^\circ$ case. On the other hand, the TE and TM gap sizes in the $60^\circ$ geometry follow the same trend as the $45^\circ$ TE case. In general, TE gaps in $35^\circ$ and TM gaps in $60^\circ$ cease to exist for PhCs with $w > 1.5a$. 

In addition to discussing the formation of photonic band gaps in photonic crystals with variation in parameters, it is imperative to evaluate the practicality and robustness of the designs in terms of fabrication. For instance, it is challenging to fabricate larger holes ($r \geq 0.43a$) in smaller widths ($w \leq 1.35a$) because of the following two issues: i) there is $\leq18\%$ (of waveguide width $w$) space  between the edge of the waveguide and the holes, and ii) only $\leq14\%$ (of unit cell $a$) room available between two adjacent holes. Therefore, a trade-off may need to be made deliberately, depending on the application, between the PBG size and the complexity of fabricating the device. Our recent work illustrates the design of a $60^\circ$ angle-etched triangular cross-section 1D PhC mirror for enhancing the quantum efficiency of \emph{in situ} superconducting nanowire single photon detectors (SNSPDs) \cite{majety2022snspd}. Even though $60^\circ$ geometry does not provide the largest complete PBG, waveguide in this geometry supports single mode propagation for NV center emission in 4H-SiC which is essential for single photon detection as well as quantum communication \cite{yusof2019ribwvg,westig2020singlemode}.   

Individual geometry offers distinct applications based on the PBG formation. We foresee three such applications from which integrated photonics with triangular geometry can benefit greatly. Operational ranges, discussed in the following, are scaled to fit wavelengths of different 4H-SiC color center emissions by updating lattice constant $a$ which depends on the choice of the center wavelength of the band gap.

\begin{figure}[ht!]
\centering\includegraphics [width=\linewidth]{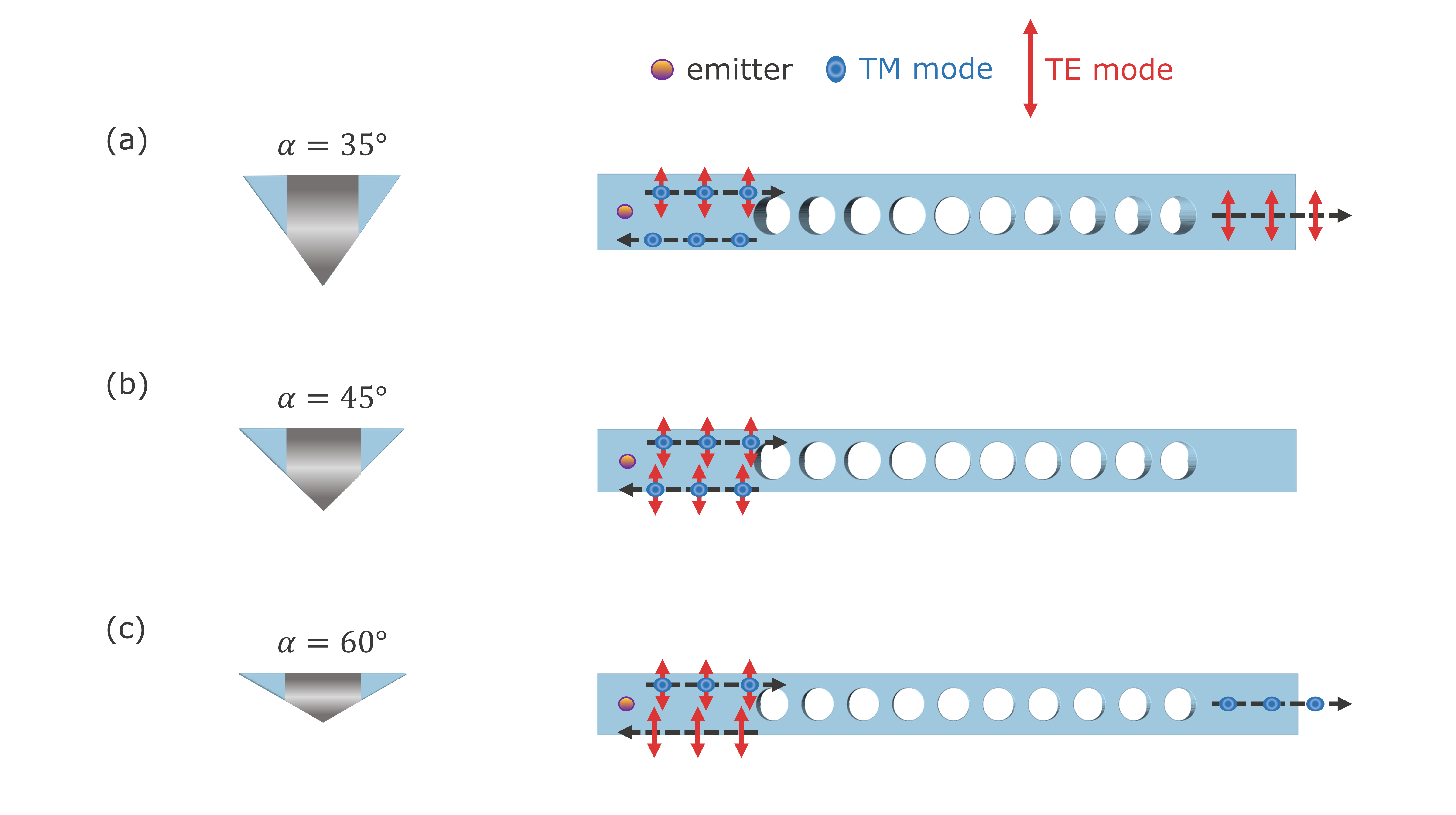}
\caption{(a) TE-pass filter in $35^\circ$ waveguide with $(a, w, r) = (390, 682, 156)$ nm. (b) Polarization-independent mirror in $45^\circ$ waveguide with $(a, w, r) = (475, 570, 190)$ nm. (c) TM-pass filter in $60^\circ$ waveguide with $(a, w, r) = (337, 590, 135)$ nm. The dashed (black) arrow shows the direction of light propagation.}
\label{fig:fig_5}
\end{figure}

Quantum communication through optical fiber network requires emission around the telecommunication bands for minimal loss of information. Vanadium ($\mathrm{V}^{4+}$) defects in 4H-SiC covers the entire O-band spectrum with 25-50\% Debye-Waller factor \cite{wolfowicz2020vanadium}. The photoluminescence features show that $\mathrm{V}^{4+}$ in 4H-SiC mostly emits TE-polarized light \cite{spindlberger2019optical}. One can make a TE-pass filter (Figure \ref{fig:fig_5}a) for vanadium color center emission in the 1285 - 1344 nm range where the TM band gap forms in the $35^\circ$ waveguide with $(a, w, r) = (390, 682, 156)$ nm photonic crystal parameters.

Silicon ($\mathrm{V}_\mathrm{Si}$) vacancy in 4H-SiC exhibits excellent optical stability and coherent spin control even in triangular waveguide structures with emissions from 861 nm to 918 nm \cite{widmann2015coherent,nagy2019high,babin2022fabrication}. The dipole polarization of single silicon vacancy center in 4H-SiC is mostly TM \cite{radulaski2017scalable} and $60^\circ$ waveguide with $(a, w, r) = (337, 590, 135)$ nm operates as a TM-pass filter (Figure \ref{fig:fig_5}c) for $\mathrm{V}_\mathrm{Si}$ emission from 840 nm to 1015 nm where the TE band gap is formed.

Longer spin coherence time is necessary for photonic-cluster based quantum computation and communication. With the combination of isotopic purification and dynamic decoupling, spin coherence time of 5s has been reported in neutral divacancy ($\mathrm{VV}^0$) in SiC\cite{anderson2022five} which is the highest among the defect spin qubits in SiC. Due to the nature of electronic structure, selection rules, and symmetry\cite{economou2016spin,wang2020NVSiC}, divacancy emission has both TE and TM polarizations  with zero-phonon line ranging from 1078 nm to 1132 nm. A polarization-independent mirror (Figure \ref{fig:fig_5}b) for divacancy emission can be constructed using the $45^\circ$ waveguide with the parameters $(a, w, r) = (475, 570, 190)$ nm that provides complete PBG from 1087 nm to 1132 nm. This happens because of the large overlapping TE-TM band gap region formed in the $\alpha = 45^\circ$ triangular cross-section waveguide PhC.


\section{Conclusion}
Photonic integration of SiC color centers is known to enhance the efficiency of quantum hardware \cite{castelletto2022silicon}. Here, the triangular geometry of devices can provide a combination of the pristine optical properties of the implanted color centers and the sample-agnostic nanofabrication. We have presented how photonic band gaps can be formed and applied in this geometry. As color centers have optical dipole-like emissions, the exploration of the dispersion relations and PBG formations in triangular cross-section 1D PhCs can play a significant role for robust light confinement in quantum photonic hardware. Our simulated results show that the nature of PBG configurations primarily depends on the etch-angle and varies intuitively with other PhC parameters. The three proposed devices can control quantum light propagation with mode selectivity and the parameter features are suitable for nanofabrication as well. These designs have the potential to improve the performance of integrated photonic devices with applications in quantum communication and quantum computing.

\section*{Acknowledgements}
This work is supported by the National Science Foundation (CAREER-2047564).

\section*{Disclosures}
The authors declare no conflicts of interest.

\section*{Data Availability Statement}
The data that support the findings of this study are available upon reasonable request from the corresponding author Pranta Saha (\textcolor{blue}{\url{prsaha@ucdavis.edu}})

\bibliographystyle{unsrt}
\bibliography{References.bib}
\end{document}